# Anomalous behaviour of the temperature dependencies of the upper critical fields in $(Dy_{1-x}Er_x)Rh_{3.8}Ru_{0.2}B_4$ ($x$=0, 0.2, 0.4).

A.V. Terekhov[1], A.P. Kazakov[2], P.M. Fesenko[3], V.M. Yarovyi[1], I.V. Zolochevskii[1], L.O. Ishchenko[1]

[1] *B. Verkin Institute for Low Temperature Physics and Engineering of the National Academy of Sciences of Ukraine, 47 Nauky Ave., Kharkiv, Ukraine, 61103*

[2] *International Research Centre MagTop, Institute of Physics, Polish Academy of Sciences, Al. Lotników 32/46, PL-02-668 Warsaw, Poland*

[3] *National Technical University «Kharkiv Polytechnic Institute», 2 Kyrpychova str., Kharkiv, Ukraine, 61002*

e-mail: terekhov@ilt.kharkiv.ua

## Introduction

One of the key areas in modern solid-state physics is the search for triplet superconductors, where the electrons in a Cooper pair have the same spin orientation. It has been shown that, in such superconductors, the pairing mechanism is mediated by spin fluctuations rather than phonons, as in conventional superconductors. It was also established that such superconductivity is very likely to be observed in magnetic superconductors, where long-range magnetic order and superconductivity coexist. It should be noted that triplet superconductors are expected to be widely used in spintronics in the future, as the electric current in such materials not only flows without scattering but is also fully spin-polarized below the transition to the superconducting state. Besides, triplet superconductivity is a necessary component for the artificial creation of Majorana states, which would be stable for a long time and resistant to external influences, and based on which qubits for quantum computers are expected to be created [1,2].

The coexistence of antiferromagnetism and superconductivity is most common. In this case, the moments of the two magnetic sublattices are compensated so that the total magnetization is zero or close to it, and the effect of magnetism on superconductivity is weak. In such materials, antiferromagnetic ordering occurs below the transition to the superconducting state. In this case, superconductivity can be partially suppressed, and the two types of ordering can coexist [3]. What would happen in the case of ferromagnetic ordering? In this case, the superconductivity is destroyed or strongly suppressed during the transition to the ferromagnetic state. For example, in the compound $ErRh_4B_4$, there is a transition to the superconducting state at 8.7 K, and then ferromagnetism appears at 0.9 K, suppressing the superconductivity [3]. However, in a narrow temperature range (before the transition to the ferromagnetic state), superconductivity and magnetism can coexist in this compound in the form of an inhomogeneous magnetic structure with sinusoidal modulation of the magnetization [4].

All of the above applies to singlet superconductors, where the electrons in the Cooper pairs have opposite spin orientations. In triplet superconductors, the electrons in the Cooper pairs have the same spin orientation. In addition, in such superconductors, electrons in the Cooper pair are coupled through spin fluctuations rather than phonons as in conventional ones. As the temperature decreases, such superconductors often exhibit the establishment of a ferromagnetic order and then the appearance of superconductivity coexisting with a ferromagnetic state over a wide temperature range. Examples of such materials are the heavy fermion compounds URhGe ($T_{FM} \approx 10$ K, $T_c \approx 0.3$ K) at ambient pressure [5] and $UGe_2$ ($T_{FM} \approx 10$ K, $T_c \approx 0.7$ K) at 1 GPa [6].

Convenient objects for elucidating aspects of the coexistence of superconductivity and long-range magnetic ordering are rare-earth borides of rhodium with dysprosium, which have a tetragonal volume-centered crystal structure of the $LuRu_4B_4$-type [7]. In these materials, the transition to magnetically ordered and then superconducting states is observed as the temperature decreases. It is not uncommon for magnetism to coexist with superconductivity in such materials down to the lowest temperatures [8,9], and thus, there is a high probability that triplet superconducting pairing is present.

The authors of article [8] have shown that in $Dy_{1-x}Y_xRh_4B_4$ compounds, the magnetic ordering temperature ($T_M$) decreases from 37 K in $DyRh_4B_4$ to 7 K in $Dy_{0.2}Y_{0.8}Rh_4B_4$ as the concentration of non-magnetic Y increases. In contrast, the superconducting transition temperature ($T_c$) increases with increasing Y concentration from 4,7 K for DyRh$_4$B$_4$ to 10.5 K for YRh$_4$B$_4$ [8]. Heat capacity measurements of the compounds Dy$_{0.8}$Y$_{0.2}$Rh$_4$B$_4$, Dy$_{0.6}$Y$_{0.4}$Rh$_4$B$_4$ and Dy$_{0.6}$Y$_{0.4}$Rh$_{3.85}$Ru$_{0.15}$B$_4$ indicate that another magnetic transformation takes place below the superconducting transition temperature ($T_{M2}$ = 1.5÷2.7 K) [9,10] and coexists with superconductivity without destroying it. Intriguing was the discovery in Dy$_{1-x}$Y$_x$Rh$_4$B$_4$ ($x$ = 0.2, 0.3, 0.4, 0.6 and 1) of the paramagnetic Meissner effect (Wohlleben effect): when cooled in a weak magnetic field, a jump with a positive value of the moment is observed in the temperature dependence of the magnetization below the superconducting transition temperature instead of a negative diamagnetic jump, as is typical for conventional superconductors [11,12]. In addition, in Dy$_{1-x}$Y$_x$Rh$_4$B$_4$ ($x$ = 0, 0.2, 0.4 and 1), non-monotonic behavior of the $H_{c2}(T)$ and $\Delta(T)$ dependences [9,13] and a strong dependence of $H_{c2}(T)$ and $R(T)$ on the slope of the external magnetic field [14] were discovered. All these results did not fit well into the framework of the Bardeen-Cooper-Schrieffer (BCS) theory for conventional superconductors and could be evidence for the presence of an unconventional mechanism of superconducting pairing.

It is known that partial replacement of rhodium by ruthenium in Dy$_{1-x}$Y$_x$Rh$_4$B$_4$ can lead to a change in the type of magnetic interactions in the Dy subsystem due to a change in the electronic structure and, as a consequence, the RKKY-exchange interaction, which determines the type of magnetic ordering [7,15]. As a result of such a replacement, the superconducting transition temperature also changes. The study of the superconductor Dy$_{0.6}$Y$_{0.4}$Rh$_{3.85}$Ru$_{0.15}$B$_4$ [16] has shown that this compound exhibits ferromagnetic ordering, which coexists with superconductivity at low temperatures. The stimulation of superconductivity (superconducting gap - $\varDelta$) by an external magnetic field and the value of the ratio $2\Delta/k_\mathrm{B}T_c > 4$, which exceeds the value of 3.52 for conventional singlet superconductors, were detected by Point-contact Andreev reflection spectroscopy [17].

It would be interesting to investigate the effect of superconductivity on rare earth rhodium borides, replacing one magnetic rare earth with another. For this purpose, the replacement of europium by dysprosium in the $(Dy_{1-x}Er_x)Rh_{3.8}Ru_{0.2}B_4$ ($x = 0, 0.2, 0.4$) was investigated [18]. Free ions Dy and Er are ferromagnetic at low temperatures ($T_M \approx 85$ K for Dy and $T_M \approx 53$ K for Er). The magnetic moment values were found to be about 9.5÷10.6 $\mu_B$ for Dy and 8.3÷9 $\mu_B$ for Er. It was assumed that the resulting changes in the magnetic subsystem should also affect the superconductivity as well, which is to be investigated in this work. It was pointed out that in rare-earth rhodium borides $(Dy_{1-x}Er_x)Rh_{3.8}Ru_{0.2}B_4$, the superconducting ordering temperature $T_c^{onset}$ increases from 3.7 K for $x = 0$ to 6 K for $x = 0.4$ with increasing Er content. The suppression of the superconducting state by the magnetism of rare-earth elements was not observed down to the lowest temperature available in the experiment – 1.5 K. This may also indicate the possibility of an unconventional superconducting coupling mechanism less affected by magnetic ordering than the conventional BCS mechanism.

The aim of this work was to study in detail the behavior of the temperature dependences of the upper critical field in $(Dy_{1-x}Er_x)Rh_{3.8}Ru_{0.2}B_4$ ($x = 0, 0.2, 0.4$) compounds and describe unusual phenomena that have been observed and may indicate the presence of unusual superconductivity.

## Samples and experimental methods

Samples of $(Dy_{1-x}Er_x)Rh_{3.8}Ru_{0.2}B_4$ ($x = 0, 0.2, 0.4$) were prepared by argon-arc melting of initial components. Partial replacement of Rh by Ru allowed obtaining a sample with the tetragonal body-centered crystalline structure of $LuRu_4B_4$-type [7] at normal pressure, which would not have been feasible without such a replacement. The obtained polycrystals were annealed for 2 days at 800°C. As a result, polycrystalline samples with tetragonal body-centered perovskite-type crystal structure $LuRu_4B_4$ (space group I4/mmm) were obtained, as evidenced by the results of *X*-ray diffraction analysis.

The resistive measurements were performed using a standard four-probe circuit on an automated physical property measurement system (PPMS) from Quantum

Design. 25 μm gold wires with silver paint served as current and potential contacts. The measurements were carried out with alternating current ($I = 100$ μA, $f = 178$ Hz), directed along the larger sample size in the temperature range of $2.5 \div 7$ K. Samples were measured with external electronics (SRS lock-in's 2124 and 124) through transformers.

## Experimental results and their discussion

Fig. 1 shows the temperature dependencies of the electrical resistivity of $(Dy_{0.6}Er_{0.4})Rh_{3.8}Ru_{0.2}B_4$ (Fig. 1a), $(Dy_{0.8}Er_{0.2})Rh_{3.8}Ru_{0.2}B_4$ (Fig. 1b) and $DyRh_{3.8}Ru_{0.2}B_4$ (Fig. 1c) in magnetic fields $0 \div 17$ kOe in relative units $R / R(7 K)$. This was done to make it easier to compare the temperature dependencies of electrical resistivity for different compounds. Due to the complex geometry of the samples, it was not possible to calculate their resistivity. At the same time, the absolute values of the electrical resistivity are not important in these measurements since this work is mainly focused on studying the temperature dependences of the upper critical fields. The dependences of $R / R(7 K)$ without field and in small magnetic fields (up to 6 kOe) have a form typical of the transition to the superconducting state (a sharp drop in electrical resistivity below $T_c$ and its disappearance at lower temperatures). In zero magnetic field, $T_c \approx 5.6$, 5 and 3.6 K for $(Dy_{0.6}Er_{0.4})Rh_{3.8}Ru_{0.2}B_4$, $(Dy_{0.8}Er_{0.2})Rh_{3.8}Ru_{0.2}B_4$ and $DyRh_{3.8}Ru_{0.2}B_4$, respectively. That is, the value of the superconducting transition temperature decreases with increasing Dy content and correspondingly decreasing Er. In fields above 6 kOe, a significant broadening of the superconducting transition with temperature is observed. This broadening increases with increasing europium concentration. It is possible that this is due to the fact that as the temperature drops below 2.5 K, a magnetic transition associated with the rear-earth magnetic subsystem appears, which crushes superconductivity. We plan to carry out detailed measurements below 2.5 K in subsequent experiments by measuring the heat capacity in the temperature range of $0.4 \div 7$ K to test the existence of low-temperature magnetic ordering. At the same time, the behaviour of the temperature dependences of the electrical resistivity described above can also be related to the presence of a non-traditional non-phonon mechanism of superconductivity, such as electron pairing via spin fluctuations. In order to better study the features of the superconducting state in this class of borides, the dependences of the upper critical field on temperature are plotted below and treated in the framework of the Werthamer-Helfand-Hohenberg theory (WHH) [19].

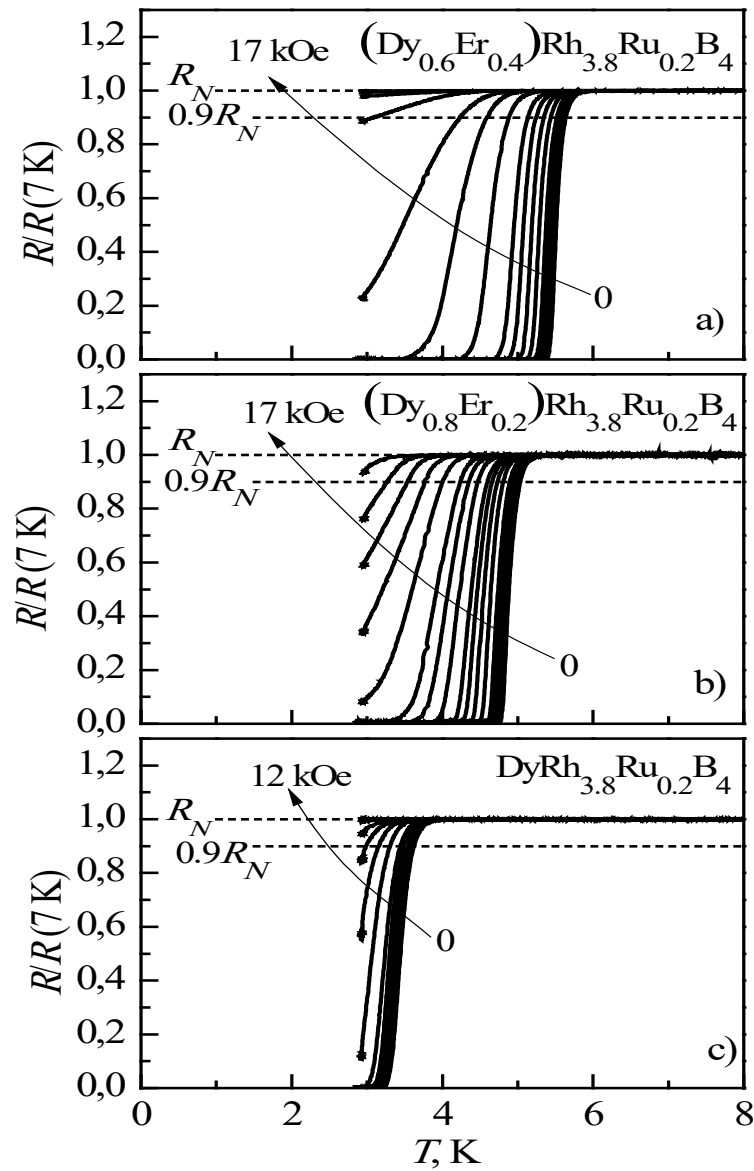

Fig. 1. Temperature dependences of electrical resistivity in relative units for borides $(Dy_{0.6}Er_{0.4})Rh_{3.8}Ru_{0.2}B_4$ (Fig. 1a), $(Dy_{0.8}Er_{0.2})Rh_{3.8}Ru_{0.2}B_4$ (Fig. 1b) and $DyRh_{3.8}Ru_{0.2}B_4$ (Fig. 1c) in magnetic fields up to 17 kOe.

Using the data of Fig. 1 and accepting for $H_{c2}(T)$ the values of the external magnetic field and the temperature, at which $R/R(7\text{ K})(H, T) = 0.9\, R_N$ ($R_N$ is the sample resistance in normal state, in relative units $R/R(7\text{ K}) = 1$), we plotted the experimental temperature dependencies of the upper critical field (Fig. 2).

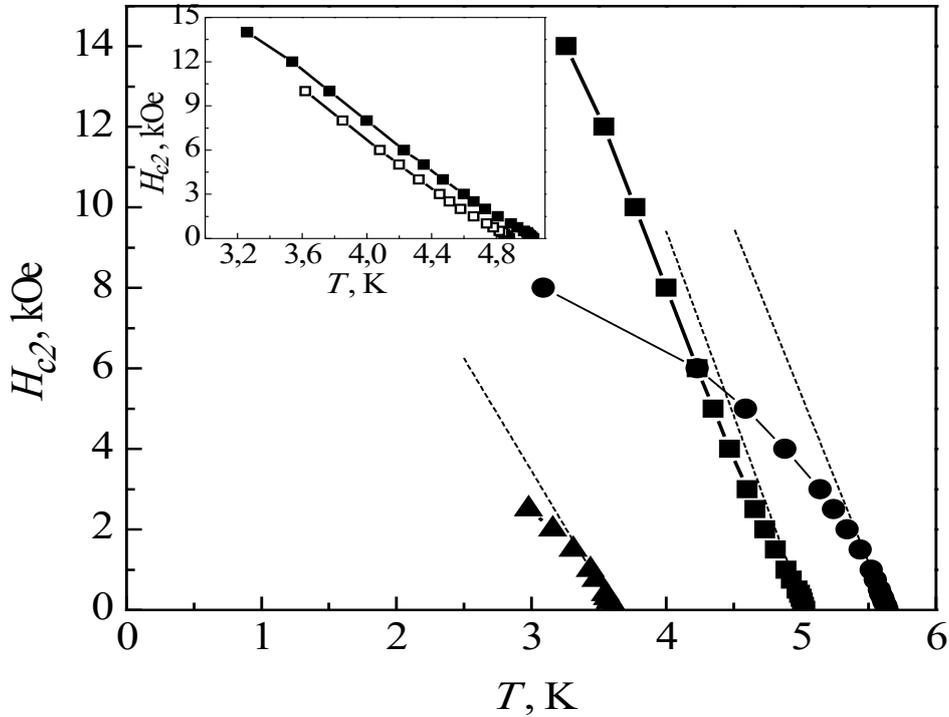

Fig. 2. Temperature dependences of the upper critical fields $H_{c2}$ for $(Dy_{0.6}Er_{0.4})Rh_{3.8}Ru_{0.2}B_4$ (●), $(Dy_{0.8}Er_{0.2})Rh_{3.8}Ru_{0.2}B_4$ (■) and $DyRh_{3.8}Ru_{0.2}B_4$ (▲). Dashed lines show linear approximation. The inset in Fig. 2 shows the $H_{c2}(T)$ dependences for $(Dy_{0.8}Er_{0.2})Rh_{3.8}Ru_{0.2}B_4$ based on the data from the $R/R(7\text{ K})(H, T)$ dependences at the level of $0.9\, R_N$ (■) and $0.5\, R_N$ (□).

In Fig. 2, the linear approximation is shown by the dashed line. The linear dependence well describes $H_{c2}(T)$ for all compounds in small magnetic fields up to 500 Oe. In the compound with an intermediate composition of Dy and Er $Dy_{0.8}Er_{0.2}Rh_{3.8}Ru_{0.2}B_4$, the experimental $H_{c2}(T)$ dependence is significantly different from those for $(Dy_{0.6}Er_{0.4})Rh_{3.8}Ru_{0.2}B_4$ and $DyRh_{3.8}Ru_{0.2}B_4$. For this composition, $H_{c2}(T)$ increases faster with decreasing temperature (superconductivity is suppressed in higher magnetic fields) than for the other two compounds, and approximately in the range 2.5 – 3.5 kOe a kink is observed that is absent for the other compositions

(Fig. 2 and inset). Is it possible that the kink in the $H_{c2}(T)$ dependence for $Dy_{0.8}Er_{0.2}Rh_{3.8}Ru_{0.2}B_4$ is due to low-temperature magnetic ordering? This explanation could take place. Superconductivity in these compounds is associated with the $4d$ electrons of the transition metal (Rh, Ru) interacting very weakly with an ordered sublattice of magnetic rare earth ions. The magnetism between these ions is due to the existence of the Ruderman-Kittel-Kasuya-Yosida (RKKY) indirect exchange interaction between rare earth ions mediated by the Rh and Ru conduction electrons. The absolute value of the exchange interaction parameter for $RERh_4B_4$ compounds is only about 0.01 eV/atom, which is at least an order of magnitude lower than for other magnetic compounds [7]. As a result, the suppression of superconductivity by magnetism may not be clearly visible in the temperature dependence of electrical resistivity. However, it can manifest itself in superconducting parameters such as the upper critical field $H_{c2}(T)$ or the superconducting gap, where features related to magnetic ordering may appear. Another possible explanation for the kink in the $H_{c2}(T)$ dependence is a transition from conventional singlet to triplet superconductivity, as theoretically predicted in [20,21].

In order to clarify the details of superconductivity in the studied rare-earth borides, we will process the experimental dependences of $H_{c2}(T)$ in the framework of the Werthamer-Helfand-Hohenberg theory (WHH). [19]. In Fig. 3, the dotted line shows the dependencies that are constructed from the calculated data within the WHH theory [19] using the expression:

$$ln\frac{1}{t} = \left(\frac{1}{2} + \frac{i\lambda_{so}}{4\gamma}\right)\psi\left(\frac{1}{2} + \frac{\bar{h} + \frac{1}{2}\lambda_{so} + i\gamma}{2t}\right) + \left(\frac{1}{2} - \frac{i\lambda_{so}}{4\gamma}\right)\psi\left(\frac{1}{2} + \frac{\bar{h} + \frac{1}{2}\lambda_{so} - i\gamma}{2t}\right) - \psi\left(\frac{1}{2}\right) \quad (1)$$

Here: $\psi$ is the digamma function, and

$$\gamma \equiv \left[\left(\alpha\bar{h}\right)^2 - \left(\frac{1}{2}\lambda_{so}\right)^2\right]^{\frac{1}{2}} \quad (2),$$

$$h^* = -\frac{H_{c2}}{(dH_{c2}/dt)|_{t=1}} = (\pi^2/4)\bar{h} \quad (3), \text{ where } t = T/T_c$$

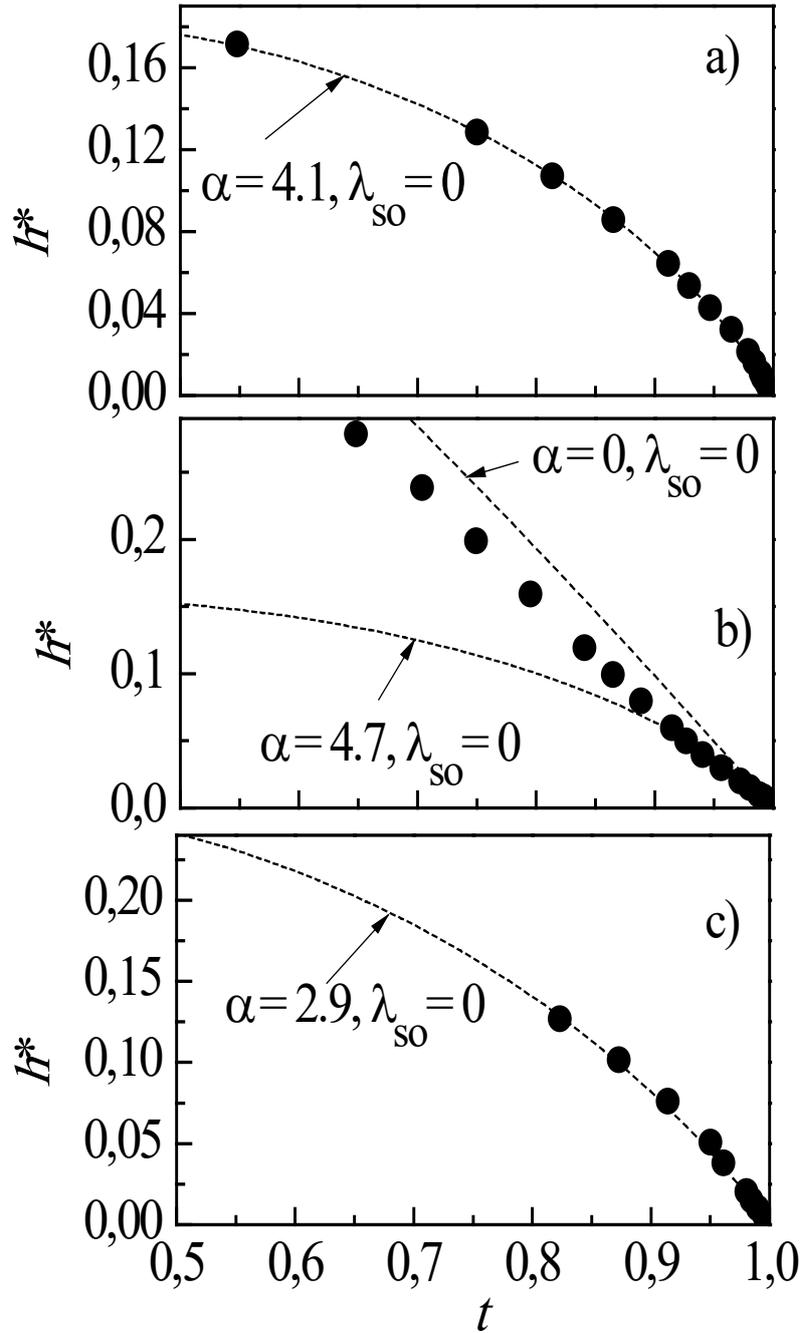

Fig. 3. The dimensionless upper critical magnetic field $h^*$ as a function of the dimensionless temperature $t = T/T_c$ for $(Dy_{0.6}Er_{0.4})Rh_{3.8}Ru_{0.2}B_4$ (a), $(Dy_{0.8}Er_{0.2})Rh_{3.8}Ru_{0.2}B_4$ (b) and $DyRh_{3.8}Ru_{0.2}B_4$ (c). The dashed curves were calculated using the WHH theory.

The fits in the calculation were: $\alpha$ – the Maki parameter, which describes the relative contribution of spin paramagnetic (a magnetic field can flip the spin of one electron in a spin-singlet Cooper pair, aligning both spins with the field direction, which breaks the pair) and orbital effects (the Lorentz force induced by the magnetic field can exceed the binding force between the electrons, causing the Cooper pair to break) in the absence of spin-orbit scattering, where $\lambda_{so}$ is parameter of spin-orbit scattering. In conventional BCS superconductors, only orbital contribution is present, i.e. $\alpha = 0$, $\lambda_{so} = 0$. The value $\alpha \neq 0$ indicates the presence of a contribution from paramagnetic effects. As $\alpha$ increases, the contribution of such effects increases accordingly. In our case, fitting the experimental values in dimensionless units $h^*(t)$ in the framework of WHH theory (by formula (1)) for all compounds gives values of the Maki parameter $\alpha > 0$, indicating a significant contribution of spin paramagnetic effects, which are usually not manifested in traditional BCS superconductors (Fig. 3). The maximum value of $\alpha = 4.7$ occurs in the compound with intermediate composition Dy - $(Dy_{0.8}Er_{0.2})Rh_{3.8}Ru_{0.2}B_4$. Correspondingly, the contribution of paramagnetic effects is largest in this compound. At the same time, the type of dependence $h^*(t)$ for $(Dy_{0.8}Er_{0.2})Rh_{3.8}Ru_{0.2}B_4$, as well as $H_{c2}(T)$, differs from similar dependencies for other compositions (Fig. 3b). The WHH theory describes well the experimental dependence only up to magnetic fields of 3 kOe (corresponding to $h^* = 0.0596$ in Fig. 3b). This may be due to low-temperature magnetic ordering in the Dy and Er subsystem. This magnetic ordering has a weaker exchange interaction (possibly antiferromagnetic) as a result of which superconductivity in a magnetic field is more weakly suppressed by paramagnetic effects. Thus, above 3 kOe, the Maki coefficient $\alpha$ can decrease significantly, as can the influence of paramagnetic effects. In Fig. 3b, for comparison, the dependence $h^*(t)$ for $\alpha = 0$ is shown, from which it is clear that, indeed, $h^*(t)$ for fields larger than 3 kOe can have a smaller coefficient $\alpha$. At the same time, an explanation related to the transition from ordinary singlet to triplet superconductivity, which was theoretically predicted in [20,21], is not excluded.

# Conclusions

For the first time, a detailed analysis of the behavior of the temperature dependences of upper critical fields has been carried out in the compounds $(Dy_{1-x}Er_x)Rh_{3.8}Ru_{0.2}B_4$ ($x$ = 0, 0.2, 0.4).

It is established that the temperature dependence of the upper critical field in the intermediate compound $(Dy_{0.8}Er_{0.2})Rh_{3.8}Ru_{0.2}B_4$ has an inflection point at 3 kOe, which may be associated with low-temperature magnetic ordering, while a more exotic mechanism caused by the transition from ordinary singlet to triplet superconductivity is not excluded.

For the first time, the experimental $H_{c2}(T)$ dependences of the $(Dy_{1-x}Er_x)Rh_{3.8}Ru_{0.2}B_4$ compounds ($x$ = 0, 0.2, 0.4) were fitted within the framework of the WHH theory. For all compounds, the Maki parameter $\alpha > 0$ indicates that spin-paramagnetic effects driven by magnetic exchange interactions play a significant role in suppressing superconductivity, which is not observed in conventional superconductors.

# Acknowledgments

The authors are grateful to Professor Yu.G. Naidyuk and Ph.D. E.Yu. Beliayev for their useful discussions and remarks, which were helpful to the process of finalizing this article. The work was supported by the National Academy of Sciences of Ukraine within the F19-5 project and Project IMPRESS-U: N2403609 (STCU project #7120); by the Foundation for Polish Science project "MagTop" no. FENG.02.01-IP.05-0028/23 co-financed by the European Union from the funds of Priority 2 of the European Funds for a Smart Economy Program 2021-2027 (FENG) and by Narodowe Centrum Nauki (NCN, National Science Centre, Poland) IMPRESS-U Project No. 2023/05/Y/ST3/00191.

Abstract

**Anomalous behaviour of the temperature dependencies of the upper critical fields in $(Dy_{1-x}Er_x)Rh_{3.8}Ru_{0.2}B_4$ ($x$=0, 0.2, 0.4).**

A.V. Terekhov, A. Kazakov, P.M. Fesenko, V.M. Yarovyi, I.V. Zolochevskii, L.O. Ishchenko

For the first time, a detailed analysis of the behaviour of the temperature dependences of the upper critical fields $H_{c2}(T)$ has been carried out in the compounds $(Dy_{1-x}Er_x)Rh_{3.8}Ru_{0.2}B_4$ ($x$ = 0, 0.2, 0.4). It is found that the $H_{c2}(T)$ in $(Dy_{0.8}Er_{0.2})Rh_{3.8}Ru_{0.2}B_4$ has an inflection point at 3 kOe, which may be related to the low-temperature magnetic ordering, while a more exotic mechanism caused by the transition from ordinary singlet to triplet superconductivity is not excluded. For the first time, the experimental $H_{c2}(T)$ dependences of $(Dy_{1-x}Er_x)Rh_{3.8}Ru_{0.2}B_4$ ($x$ = 0, 0.2, 0.4) compounds have been fitted within the framework of Werthamer-Helfand-Hohenberg theory (WHH) with the Maki parameter $\alpha > 0$, indicating that spin-paramagnetic effects due to magnetic exchange interactions play an essential role in suppressing superconductivity in these compounds.

Keywords: rare-earth rhodium borides, Werthamer-Helfand-Hohenberg theory (WHH), Maki parameter, magnetic superconductors, triplet pairing.

1. X. Liu, J. D. Sau, and S. Das Sarma, Physical Review B - Condensed Matter and Materials Physics **92**, 014513 (2015).
2. G. Livanas, M. Sigrist, and G. Varelogiannis, Scientific Reports **9**, 6259 (2019).
3. A. I. Buzdin, L. N. Bulaevskii, M. L. Kulich, and S. V. Panyukov, Uspekhi Fizicheskih Nauk **144**, 597 (1984).
4. H. A. Mook, O. A. Pringle, S. Kawarazaki, S. K. Sinha, G. W. Crabtree, D. G. Hinks, M. B. Maple, Z. Fisk, D. C. Johnston, L. D. Woolf, and H. C. Hamaker, Physica B+C **120**, 197 (1983).
5. D. Aoki, A. Huxley, E. Ressouche, D. Braithwaite, J. Flouquet, J. P. Brison, E. Lhotel, and C. Paulsen, Nature **413**, 613 (2001).
6. S. S. Saxena, P. Agarwal, K. Ahilan, F. M. Grosche, R. K. W. Haselwimmer, M. J.

Steiner, E. Pugh, I. R. Walker, S. R. Julian, P. Monthoux, G. G. Lonzarich, A. Huxley, I. Sheikin, D. Braithwaite, and J. Flouquet, Nature **406**, 587 (2000).

7. Maple M.B. and Fischer O., *Superconductivity in Ternary Compounds II, Superconductivity and Magnetism* (Springer New York, NY, New York, 1982).

8. V. M. Dmitriev, A. J. Zaleskl, E. P. Khlybov, L. F. Rybaltchenko, E. V. Khristenko, L. A. Ishchenko, and A. V. Terekhov, Acta Physica Polonica A **114**, 83 (2008).

9. V. M. Dmitriev, A. Zaleskiǐ, E. P. Khlybov, L. F. Rybal'Chenko, E. V. Khristenko, L. A. Ishchenko, A. V. Terekhov, I. E. Kostyleva, and S. A. Lachenkov, Low Temperature Physics **34**, 909 (2008).

10. A. V. Terekhov, I. V. Zolochevskii, L. A. Ishchenko, A. Zaleski, E. P. Khlybov, and S. A. Lachenkov, Low Temperature Physics **42**, 232 (2016).

11. V. M. Dmitriev, A. V. Terekhov, A. Zaleski, E. N. Khatsko, P. S. Kalinin, A. I. Rykova, A. M. Gurevich, S. A. Glagolev, E. P. Khlybov, I. E. Kostyleva, and S. A. Lachenkov, Low Temp. Phys. **38**, 154 (2012).

12. A. V. Terekhov, Low Temperature Physics **39**, 640 (2013).

13. V. M. Dmitriev, A. Zaleski, E. P. Khlybov, L. F. Rybaltchenko, E. V. Khristenko, L. A. Ishchenko, and A. V. Terekhov, Low Temperature Physics **35**, 424 (2009).

14. A. V. Terekhov, I. V. Zolochevskii, E. V. Khristenko, L. A. Ishchenko, E. V. Bezuglyi, A. Zaleski, E. P. Khlybov, and S. A. Lachenkov, Physica C: Superconductivity and Its Applications **524**, 1 (2016).

15. K. J. B. Bennemann K. H., editor, *Superconductivity: Conventional and Unconventional Superconductors* (Springer Berlin Heidelberg, Berlin, Heidelberg, 2008).

16. A. V. Terekhov, I. V. Zolochevskii, L. A. Ischenko, A. N. Bludov, A. Zaleski, E. P. Khlybov, and S. A. Lachenkov, Low Temperature Physics **45**, 1241 (2019).

17. L. F. Rybaltchenko, E. V. Khristenko, L. A. Ishchenko, A. V. Terekhov, I. V. Zolochevskii, T. V. Salenkova, E. P. Khlybov, and A. J. Zaleski, Low Temperature Physics **38**, 1106 (2012).

18. A. V. Terekhov, V. M. Yarovyi, I. V. Zolochevskii, L. O. Ishchenko, and E. V.

Khrystenko, Low Temp. Phys. **49**, 1087 (2023).

19. N. R. Werthamer, E. Helfand, and P. C. Hohenberg, Physical Review **147**, 295 (1966).

20. M. M. Maśka, M. Mierzejewski, B. Andrzejewski, M. L. Foo, R. J. Cava, and T. Klimczuk, Physical Review B - Condensed Matter and Materials Physics **70**, 144516 (2004).

21. M. Mierzejewski, M. Maśka, and B. Andrzejewski, Acta Physica Polonica A **106**, 603 (2004).